\documentclass[aps,prfluids,superscriptaddress,nofootinbib]{revtex4-2}

\usepackage{fullpage}
\usepackage{hyperref}
\hypersetup{
    colorlinks=true,
    linkcolor=blue,
    filecolor=magenta,      
    urlcolor=cyan,
    citecolor = {blue}
    }
\usepackage{amssymb,euscript,latexsym,amsmath}
\usepackage{lipsum}
\usepackage{graphicx}
\usepackage[dvips]{epsfig}
\usepackage{color}
\usepackage[dvipsnames]{xcolor}
\usepackage{epstopdf}
\usepackage{natbib}
\usepackage{caption, subcaption}
\DeclareGraphicsExtensions{.eps,.ps,.pdf,.png}

\DeclareGraphicsExtensions{.eps,.ps,.pdf}

\DeclareMathOperator{\csch}{csch}

\DeclareMathOperator{\arccsch}{arcsch}
 
\newcommand{\abs}[1]{\left|#1\right|}

\begin{document}


\title{Hydrodynamic bound states of rotating micro-cylinders in a confining geometry}

\author{Hanliang Guo}\email[]{hguo@owu.edu}
\affiliation{Department of Mathematics and Computer Science, Ohio Wesleyan University, Delaware, OH 43015, USA}
\author{Yi Man}
\affiliation{Department of Mechanics and Engineering Science at College of Engineering, and State Key Laboratory for Turbulence and Complex Systems, Peking University, Beijing 100871, China}
\author{Hai Zhu}
\affiliation{Center for Computational Mathematics, Flatiron Institute, New York, NY 10010, USA}


\date{\today}

\begin{abstract}
Many micro-swimmers propel themselves by rotating micro-cylindrical organelles such as flagella or cilia. These cylindrical organelles almost never live in free space, yet their motions in a confining geometry can be counter-intuitive. For example, one of the intriguing yet classical results in this regard is that a rotating cylinder next to a plane wall does not generate any net force in Newtonian fluids and therefore does not translate. In this work, we employ analytical and numerical tools to investigate the motions of micro-cylinders under prescribed torques in a confining geometry. We show that a cylinder pair can form four non-trivial hydrodynamic bound states depending on the relative position within the confinement. Our analysis shows that the distinct states are the results of competing effects of the hydrodynamic interactions within the cylinder pair and between the active cylinders and the confinement.
\end{abstract}


\maketitle


\section{Introduction}

Rotation is a fundamental form of microorganism locomotion. Flagellated bacteria such as {\em E. coli} move by rotating their semi-rigid flagella~\citep{berg1993random}; green algae  {\em Chlamydomonas} swim in a helical trajectory while rotating its cell-body with the beat of two near-identical flagella~\citep{ruffer1985high,schaller1997chlamydomonas}; thousands of flagella on the surface of {\em Volvox} generate a tangential velocity at an angle of the axis of motion that generates the swirling motion~\citep{kirk1998volvox}.
Even though body rotation may not be the most efficient way of locomotion through the lens of hydrodynamics, it can be biologically beneficial as it allows the microorganisms to perceive the environment from all angles and enable crucial functions such as photo-taxis~\citep{pedley1992hydrodynamic}.
From the technological point of view, micro-particles can be driven into rotation mode relatively easily through external fields such as magnetic fields, and the driven rotations are usually coupled with translations and can lead to interesting collective behaviors~\citep{tierno2021transport}. Understanding the mechanisms of these micro-rotors or micro-rollers in complex conditions has been a point of focus recently as they have shown great potential in drug delivery, microsurgery, and mixing~\citep{tierno2008controlled,sing2010controlled,alapan2020multifunctional,ahmed2021bioinspired}.

Biological microorganisms and artificial microswimmers almost never live in free space, and it has been known for many decades that nearby boundaries qualitatively alter the dynamics of microswimmers. 
For example, \citet{rothschild1963non} reported that Bull spermatozoa are attracted to no-slip boundaries, and suggested that it might be due to the hydrodynamic interaction between the spermatozoa and the boundary about 60 years ago.
Numerous studies have since revealed the effects of hydrodynamic interactions from the confinement geometry to the microswimmers, both experimentally and theoretically. 
Among many other interesting findings, we now know that {\em E. coli} swim in circular motions near the boundary due to the force- and torque-free swimming and the hydrodynamic interactions with the boundary, and the direction of the motion depends on the boundary type~\citep{lauga2006swimming,lemelle2010counterclockwise,di2011swimming};
active particles can be trapped into closed orbits by passive colloids, and can also trap and transport passive cargos~\citep{takagi2014hydrodynamic,chamolly2020irreversible,van2023simple};
geometric asymmetry of the microswimmer or the boundary can also lead to qualitatively different trajectories~\citep{spagnolie2012hydrodynamics,chaithanya2021wall}.
Readers are kindly referred to a concise review of this topic given by \citet{elgeti2016microswimmers}.

Hydrodynamic interactions between multiple microswimmers and the boundary can lead them into various interesting periodic motions, called hydrodynamic bound states.
{To date, many of the models studying microswimmers assume that the swimmers do not generate torques. For example, Crowdy and coworkers studied the dynamics of treadmilling swimmers next to no-slip boundaries of various shapes~\citep{crowdy2011treadmilling,crowdy2011hydrodynamic,davis2012stresslet}. On the other hand, interactions between rotating swimmers display beautiful and intriguing dynamics that can only be studied if the model considers net torque.} For example, a pair of {\em Volvox} ``dance'' in multiple {hydrodynamic bound states} when in close proximity to solid walls, forming the {\em waltz} and {\em minuet} bound states~\citep{drescher2009dancing}; magnetically driven micro-rollers can form various states such as ``critters'' or one-dimensional chain via sole hydrodynamic interactions~\citep{driscoll2017unstable,martinez2018emergent}. 
{The modeling aspect of these problems is usually dealt with singularity methods that treat each micro-roller as a rotlet, and image systems that account for the no-slip confinement~\citep{meleshko1996blinking,tallapragada2019chaotic,delmotte2019hydrodynamically}. Despite its cleanness and mathematical elegance, rotlet models are far-field approximations and do not capture the dynamics well when the micro-rollers are in close proximity to each other and/or the confinement.
\citet{delmotte2019hydrodynamically} adopted a rotlet model and accounted for the finite size effect by using the Rotne-Prager-Yamakawa mobility with wall corrections,} and showed that rich dynamics can be obtained for a micro-roller {pair} above a flat wall. They found that the different states of the micro-roller pair can be obtained by altering the relative strength of gravitational forces and external torques. Furthermore, the micro-roller pair would be in a stable motile orbiting mode, reminiscent of the ``critters'' state observed in micro-roller suspensions, when the relative strength is high. Studying the micro-roller pair provides us the opportunity to obtain a deep understanding of the mechanisms behind the various states of micro-roller suspensions.
However, the method illustrated in \citep{delmotte2019hydrodynamically} relies on the image systems above a no-slip planar boundary \citep{blake1974fundamental} that is difficult to extend to other types of confinement.

In this paper, we focus on the hydrodynamic bound states of two neutrally buoyant rotating cylinders inside cylindrical confinement with circular cross-sectional areas. 
The dynamics of one rotating cylinder inside the confinement is derived analytically, and the dynamics of a rotating cylinder pair is computed numerically.
We note that while it is straightforward to extend the numerical method to more complex geometry, we focus on circular cylinders to allow feasible analysis of the mechanisms.
We show that a pair of rotating circular cylinders form four non-trivial hydrodynamic bound states, resulting from the competing effects of the hydrodynamic interactions within the cylinder pair and between the active cylinders and the confinement.

\section{Model and Methods}
\begin{figure}
        \centerline{\includegraphics{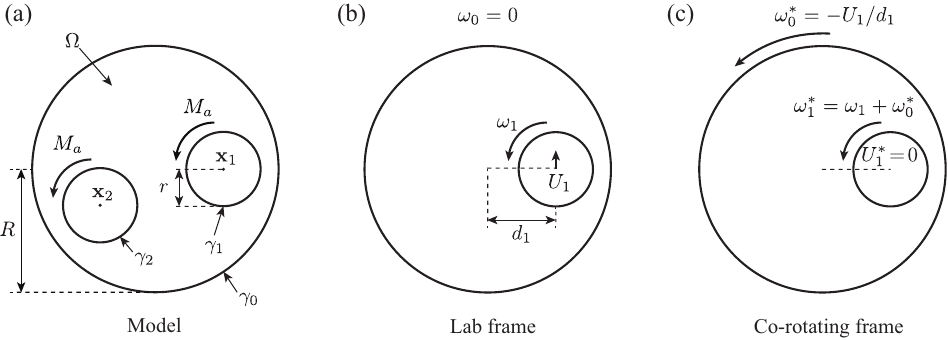}}
\caption[]{(a) Model. Active cylinders of radius $r$ inside the confining cylinder of radius $R$. The active cylinders are driven by a prescribed torque $M_a\mathbf{e}_z$ per unit length. The gap between the active cylinder and the confinement is filled with fluid of viscosity $\mu$. $\gamma_0$ denotes the surface of the confinement and $\gamma_k$ ($k>0$) denotes the surfaces of the active cylinders. $\Omega$ denotes the fluid domain bounded by the surfaces. 
(b) Single active cylinder case. The prescribed torque $M_a\mathrm{e}_z$ of the active cylinder generates an angular velocity $\omega_1\mathrm{e}_z$ and a translational velocity $U_1$ in the azimuthal direction of the confinement. The active cylinders is positioned $d_1$ units away from the center of the confinement.  
(c) Co-rotating frame that eliminates the translational velocity of the active cylinder. {In the co-rotating frame, the active cylinder is effectively ``pinned'' at the initial position, and the confinement would rotate clockwise.} Quantities with asterisks are those measured in the co-rotating frame. Positive directions of measured quantities are denoted by arrows. 
}
	\label{fig:model}
\end{figure}

Consider a cylindrical confinement $\gamma_0$ with a circular cross-section of radius $R$ filled with viscous fluids. 
Active cylinders $\gamma_k$ ($k>0$) of radius $r$ are suspended inside the confinement. Each active cylinder generates a torque per unit length $M_a\mathbf{e}_3$ about its own axis. 
Let the center of the confinement be the origin and the center of the $k$-th active cylinder be $\mathbf{x}_{k}$. The schematic figure is shown in Figure~\ref{fig:model}(a).

In the viscosity-dominant regime, the inertia is negligible and 
the fluid dynamics is governed by the incompressible {Stokes equation}:
\begin{equation}\label{eq:stokes}
-\mu \nabla^2\mathbf{u} + \nabla p = \mathbf{0}, \quad \nabla \cdot \mathbf{u} = 0, \quad \text{for }\mathbf{x}\in\Omega,
\end{equation}
where $\Omega$ is the fluid domain, $\mathbf{u}$ is the fluid velocity, $p$ is the pressure, and $\mu$ is the fluid viscosity.
The fluid velocity on the surface of the active cylinder and the confining geometry are given by rigid-body motion and the no-slip boundary condition:
\begin{equation}\label{eq:bc}
\begin{split}
\mathbf{u}(\mathbf{x}) = \left\{ \begin{array}{ll}
0, & \text{for } \mathbf{x} \in \gamma_0\\
\mathbf{U}_{k} + \omega_k\mathbf{e}_z\times(\mathbf{x} - \mathbf{x}_{k}), & \text{for } \mathbf{x}\in  \gamma_k \quad(k>0)
\end{array}\right.,
\end{split}
\end{equation}
where $\mathbf{U}_{k}$ and $\omega_k$ are the $k$-th active cylinder's centroidal translational and angular velocities respectively.
The active cylinders also satisfy the force- and torque-balance conditions in the viscosity-dominant regime:
\begin{equation}\label{moment}
\int_{\gamma_k} \mathbf{f}(\mathbf{x}) d\mathbf{x} = \mathbf0, \quad \int_{\gamma_k} (\mathbf{x} - \mathbf{x}_{k})\times \mathbf{f}(\mathbf{x}) d\mathbf{x} + M_a\mathbf{e}_z= \mathbf0,
\end{equation}
where $\mathbf{f}$ is the fluid traction density per unit length on the cylinder boundary ($k>0$).

In general, given the centroidal positions of the active cylinders, we solve $\mathbf{U}_k$ and $\omega_k$  numerically using a high-order boundary-integral method, similar to our previous work~\citep{guo2020simulating}. The details of our implementation are included in the Supplemental Material~\cite{supp}.

We note that the case of a single active cylinder rotating inside the confinement can be solved analytically. 
Specifically, we are interested in the rotation-induced translational velocity of the inner cylinder when its center is $d_1$ distance away from the center of the confining cylinder (Fig.~\ref{fig:model}(b)).
Without loss of generality, we assume the center of the active cylinder $\mathbf{x}_1$ is on the positive $x$-axis.
In this set-up, the $x$-component of the translational velocity is zero by symmetry. Thus $\mathbf{U}_1 = U_1\mathbf{e}_y$, where $U_1$ is the centroidal translational speed.

To find $U_1$, 
we adopt the bipolar coordinate systems~\citep[Appendix-12]{happel2012low} following conventional approaches (e.g., in \citep{jeffery1922rotation,jeffrey1981slow,wakiya1975application}). 
The transformation between Cartesian and the bipolar coordinate systems are given by
\begin{equation}
\left\{
\begin{array}{l}
\displaystyle x + iy = ic \cot\frac{\xi + i \eta}{2}, \quad  (c>0)\\
\displaystyle x = \frac{c\sinh \eta}{\cosh\eta - \cos\xi}, \quad  y = \frac{c\sin\xi}{\cosh\eta - \cos\xi},  
\end{array}
\right.
\end{equation}
where $\xi\in[0,2\pi]$ and $\eta\in(-\infty, \infty)$ are the bipolar coordinates, and $i=\sqrt{-1}$ is the imaginary unit.
The curves given by $\eta = \eta_0$ are a family of circles with centers at $(x,y) = (c\coth\eta_0, 0)$ and radius $c |\csch\eta_0|$. Let the active cylinder and the confinement be denoted by $\eta = \alpha$ and $\eta=\beta$ respectively, where $\alpha<\beta<0$. 
The parameters $\alpha$, $\beta$, and $c$ can be solved directly from the cylinder diameters $R$ and $r$, and the center-to-center distance $d_1$:
\begin{equation}
c = \sqrt{\left(\frac{d_1^2+r^2-R^2}{2d_1}\right)^2 - r^2}, \quad \alpha = -\arccsch (r/c), \quad \beta = -\arccsch(R/c).
\end{equation}

\citet{wakiya1975application} solved the force and moment exerted upon the cylinders when both cylinders are pinned at their centers and rotate with given angular velocities.
To leverage their solution, we apply a frame transformation to cancel the active cylinder's translational velocity $U_1\mathbf{e}_y$. 
Specifically, we choose a co-rotating frame at an angular velocity $U_1/d_1$ in the counter-clockwise direction (Fig.~\ref{fig:model}(c)). In this frame, the confining cylinder experiences an angular velocity $\omega_0^* = -U_1/d_1$, and the inner cylinder experiences an angular velocity $\omega_1^* = \omega_1 + \omega_0^*$ and zero translational velocity. Note that we are using the asterisk ($*$) to denote the quantities in the co-rotating frame.
Substituting into equations (2.13 - 2.14) in \citet{wakiya1975application} yields the force and moment exerted upon the active cylinder:
\begin{equation}
\left\{
\begin{array}{l}
F_x = 0\\
F_y = 4\pi\mu(r\omega_1^*\sinh\beta + R\omega_0^*\sinh\alpha) \sin(\alpha - \beta)/S\\
\displaystyle M = -4\pi\mu r^2\sinh^2\alpha\left\{ \omega_1^*\frac{\sinh^2(\alpha - \beta)}{\sinh^2\alpha} + (\omega_1^* - \omega_0^*)\left[(\alpha - \beta)\frac{\cosh(\alpha - \beta)}{\sinh(\alpha - \beta)}-1\right]\right\}/S
\end{array}
\right.
\end{equation}
where $S = (\alpha - \beta)(\sinh^2\alpha + \sinh^2\beta) - 2\sinh\alpha\sinh\beta\sinh(\alpha - \beta)$.
$F_x = 0$ confirms our previous argument that the active cylinder does not have radial velocity.
Note that the positive sign is used in the expression of $F_y$ as $\alpha < 0$ in our case.

Substituting the force-balance condition ($F_y = 0$) and torque-balance condition ($M + M_a = 0$) yields 
\begin{equation}
\omega_0^* = \frac{SM_a}{4\pi\mu r^2\sinh^2\alpha}\left/\left\{ \frac{R\sinh\alpha\sinh^2(\alpha-\beta)}{r\sinh\beta\sinh^2\alpha} + \left(\frac{R\sinh\alpha}{r\sinh\beta}+1\right)\left[(\alpha - \beta)\frac{\cosh(\alpha-\beta)}{\sinh(\alpha-\beta)}-1\right] \right\}\right.,
\end{equation}
and the translation speed is simply 
\begin{equation}\label{eq:trans_vel}
U_1 = -\omega_0^*d_1.
\end{equation}

\section{Results}

\begin{figure}
        \centering
        \includegraphics{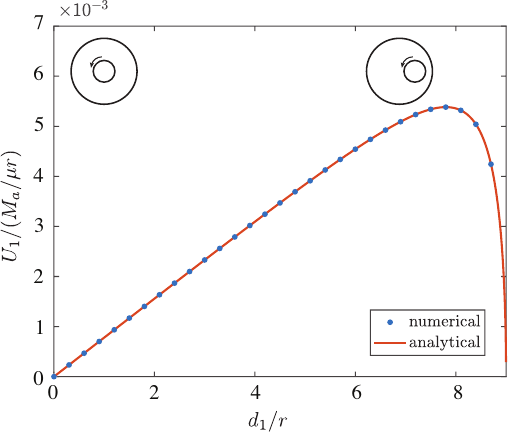}
	\caption[]{Translational velocity of the active cylinder as a function of the center position, scaled by the characteristic speed and length respectively. Numerical results are shown in blue circles, and analytical solutions are shown in the red curve. $R = 10r$. Insets denote the concentric configuration ($d_1=0$) and the non-concentric configuration ($d_1>0$). }
	\label{fig:trans_vel}
\end{figure}

Figure~\ref{fig:trans_vel} shows the translational velocity as a function of the center-to-center distance between the active cylinder and the confinement for $R=10r$. 
The analytical and numerical solutions match perfectly. 
The translational velocities at $d_1=0$ and $d_1=R-r$ are both 0, as dictated by symmetry and the no-slip boundary conditions, respectively. 
The active cylinder translates the fastest when $d_1 \approx 8r$, in which case the gap between the active cylinder and the confinement is approximately the radius of the active cylinder.
Interestingly, the translational velocity increases almost linearly for  $0\le d_1\le 6r$. This means that the orbital angular velocity of the active cylinder ($U_1/d_1$) is almost constant when $d_1 \le 6r$. In other words, if more than one active cylinders is in the $d_k \le 6r$ region and {\em not} interacting hydrodynamically with each other, the relative positions between each active cylinder will remain the same.

\begin{figure}
\centering
\includegraphics{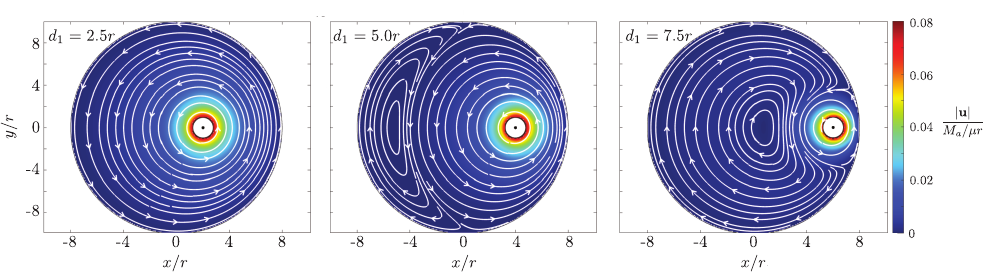}
	\caption[]{Flow fields in the lab frame for single active cylinder inside the confining cylinder  ($R/r=10$) at different positions. From left to right: $d_1/r = 2.5$, $5$, and $7.5$ respectively. The streamlines are shown on top of the flow field which is color-coded by the magnitude of the fluid velocity.}
	\label{fig:single_solution}
\end{figure}

The numerical flow fields are shown in Figure~\ref{fig:single_solution} with increasing eccentricity resulting from the position of the active cylinder ($d_1/r = 2.5,\, 5,\, 7.5$). 
When the eccentricity is high, we reproduce the classical counter-rotating secondary recirculation zone inside the confinement.
{The existence of the secondary recirculation zone, as will be shown in the following sections,  strongly affects the dynamics of the cylinder pair.}

Next, we investigate the long-term dynamics of an active cylinder pair inside the confinement. 
Specifically, consider two active cylinders 
suspended inside a stationary confining circular cylinder of radius $R = 10r$
with the same torque per unit length $M_a\mathbf{e}_z$.
No analytical solution {seems feasible} for this case and we adopt the numerical route.

\begin{figure}
\centering
\includegraphics{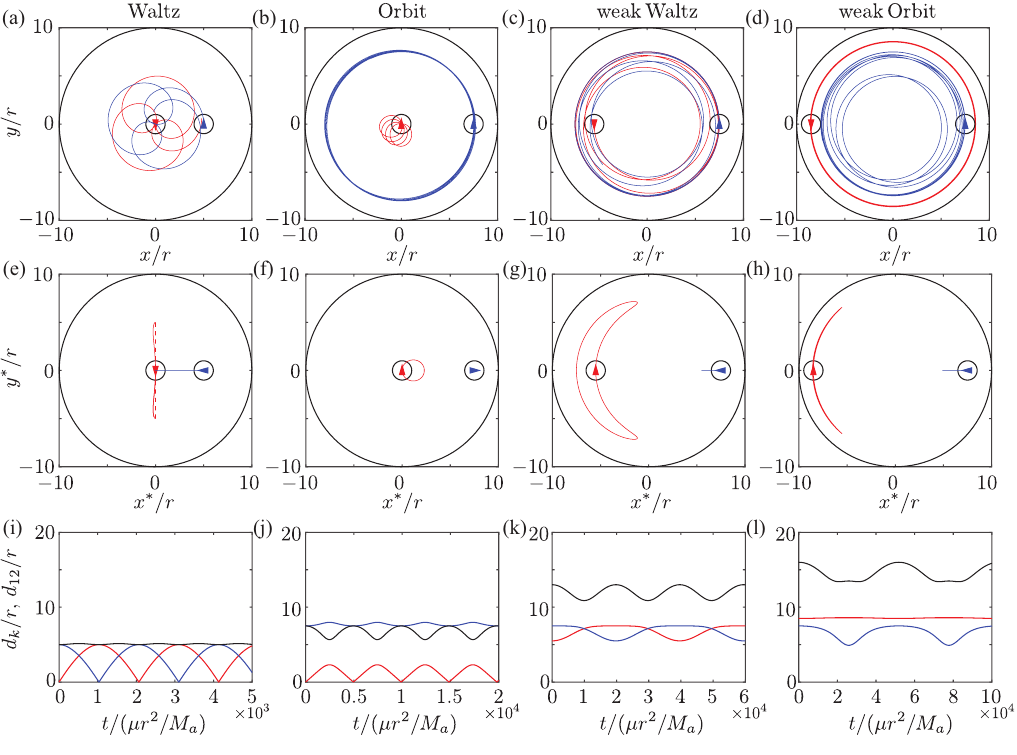}
	\caption[]{Characteristic trajectories of cylinder pairs with different initial positions.  
 (a-d): center trajectories in the lab frame; 
 (e-h): center trajectories co-rotating frame such that cylinder 1 is always on the positive $x^*$ axis; 
 (i-l): distance to the center of confinement (blue/red) and the center-to-center distance within the cylinder pair (black) as functions of time. 
 In all panels, blue and red curves are the corresponding curves for cylinder 1 and 2, respectively. The arrows in panels (a-h) denote the translational velocity at the initial positions.
 Four different hydrodynamic bound states are shown in columns.
 Waltz: $\mathbf{x}_1(0) = 5r\mathbf{e}_x, \mathbf{x}_2(0) = 0$;
 Orbit: $\mathbf{x}_1(0) = 7.5r\mathbf{e}_x, \mathbf{x}_2(0) = 0$;
 weak Waltz: $\mathbf{x}_1(0) = 7.5r\mathbf{e}_x, \mathbf{x}_2(0) = -5.5r\mathbf{e}_x$;
 weak Orbit: $\mathbf{x}_1(0) = 7.5r\mathbf{e}_x, \mathbf{x}_2(0) = -8.5r\mathbf{e}_x$.
 }
	\label{fig:trajectories}
\end{figure}

Let $\mathbf{x}_k(t)$ denote the $k$-th active cylinder's center position at time $t$ and $d_k(t) = |\mathbf{x}_k(t)|$. 
Figure~\ref{fig:trajectories} shows a gallery of four qualitatively different periodic trajectories with different initial positions.

We start by putting cylinder 1 halfway between the center and the boundary of the confinement and cylinder 2 at the center of the confinement ($\mathbf{x}_1(0) = 5.0r\mathbf{e}_x$, $\mathbf{x}_2(0) = 0$).
Both cylinders orbit counter-clockwise when viewed in the lab frame and the center of each cylinder traces a {\em cycloid-type curve} around the origin. (Fig.~\ref{fig:trajectories}(a)).
The cycloid-type curves collapse into simple curves when viewed in the co-rotating frame $x^*y^*$, implying that the motions of the cylinders are periodic (Fig.~\ref{fig:trajectories}(e)).
The co-rotating frame is defined in the same way as in our analytical method to eliminate cylinder 1's azimuthal movement.
The center of cylinder 2 traces a slightly curved trajectory approximately aligned with the $y^*$ axis. The discontinuity of cylinder 2's trajectory, denoted by the red dashed line in the co-rotating frame, happens in the degenerate case where cylinder 1 is at the center of the confinement.
The periodic nature of the motions is also reflected when we plot the distances between the active cylinders and the center of the confinement as functions of time (Fig.~\ref{fig:trajectories}(i)). 
As time increases, $d_1(t)$ first decreases from $5r$ to 0 as cylinder 1 gradually moves  towards the origin while $d_2(t)$ increases from $0$ to $5r$. At about $t = 10^3\mu r^2/M_a$, $d_1(t) = 0$ as cylinder 1 reaches the origin and $d_2(t) = 5r$. At this point, the two cylinders essentially exchanged their positions compared to their initial positions, albeit with a rotated viewing angle. 
As a result, the functions $d_1(t)$ and $d_2(t)$ are identical up to a half-period phase lag.
That is, $d_1(t) = d_2(t + T/2)$ for all $t$, where $T$ is the period of the motion.
We refer to this type of periodic motion with exchangeability between cylinders as the {\em Waltz} bound state, motivated by the dancing motion of {\em Volvox} \cite{drescher2009dancing}.

Different types of long-term dynamics can be obtained by varying the initial positions of the cylinder pair. 
For example, if we keep the initial position of cylinder 2 to be at the origin and move cylinder 1 further away such that $\mathbf{x}_1(0) = 7.5r\mathbf{e}_x$, 
 cylinder 1 will always orbit around the origin at a large yet approximately constant distance, whereas cylinder 2 is ``trapped'' to move in small circles close to the origin (Fig.~\ref{fig:trajectories}(b)). Again, the nature of the motion is best shown in the co-rotating frame: cylinder 2 orbits in a small circle biased towards cylinder 1 while cylinder 1 barely moves (Fig.~\ref{fig:trajectories}(f)).
During the period, $d_1(t)$ varies between $7.5r$ and $8.0r$ while $d_2(t)$ varies between $0$ and $2.3r$. 
The center-to-center distance of the cylinder pair is also larger compared to the first case, where $5.7r < d_{12}(t) < 7.5r$ (Fig.~\ref{fig:trajectories}(j)).
Unlike the Waltz bound state, the exchangeability within the cylinder pair is lost as $d_1(t) > d_2(t)$ for all $t$, and the two cylinders seem to orbit with their own radius. We refer to this type of period motion as the {\em Orbit} bound state.

Another two transitions in states happen as we keep increasing the center-to-center distance within the cylinder pair.
Specifically, for $\mathbf{x}_1(0) = 7.5r\mathbf{e}_x$, and $\mathbf{x}_2(0) = -5.5r\mathbf{e}_x$, no cylinder would be trapped close to the origin, and the respective distances to the confinement center are similar for the two cylinders (Fig.~\ref{fig:trajectories}(c)).
Interestingly, the center of cylinder 2 in the co-rotating frame traces a crescent shape that encloses the trajectory of cylinder 1 if mirrored about the $y^*$ axis (Fig.~\ref{fig:trajectories}(g)). 
This is a state reminiscent of the Waltz bound state as $d_1(t)$ and $d_2(t)$ are identical up to a half-period phase lag (Fig.~\ref{fig:trajectories}(k)). The larger center-to-center distance $d_{12}$ leads to weaker hydrodynamic interactions within the cylinder pair and the period is much longer compared to Waltz.
We refer to this state as the {\em weak Waltz} bound state.
On the other hand, increasing $d_{12}(0)$ further would reveal a state similar to the Orbit bound state (Fig.~\ref{fig:trajectories}(d)(h)(l)). 
Specifically, if $\mathbf{x}_1(0) = 7.5r\mathbf{e}_x$ and $\mathbf{x}_2(0) = -8.5r\mathbf{e}_x$, $d_1(t)$ oscillates between $4.9r$ and $7.5r$, while $d_2(t)$ is almost constant at $8.5r$. $d_{12}$ in this case oscillates between $13.4r$ and $16r$. Unlike the typical Orbit bound state, no cylinder is trapped close to the origin. We refer to this state as {\em weak Orbit} bound state.

We then conduct a systematic study of the initial positions of both active cylinders $\mathbf{x}_k(0)$.
Due to rotational symmetry, we set $\mathbf{x}_1(0) = d_1(0)\mathbf{e}_x$, where $0\le d_1(0) < 9r$ without loss of generality.
Additionally, numerical evidence (not shown here) suggests that no matter where $\mathbf{x}_2(0)$ is, there is always a time ${t}_o$ such that the origin, $\mathbf{x}_1({t}_o)$, and $\mathbf{x}_2({t}_o)$ form a straight line for $R=10r$. 
This result greatly reduces the parameter space we need to explore, as we only need to consider the cases where $\mathbf{x}_2(0)$ is on the $x$-axis as well.

\begin{figure}
\centering
\includegraphics[width=\linewidth]{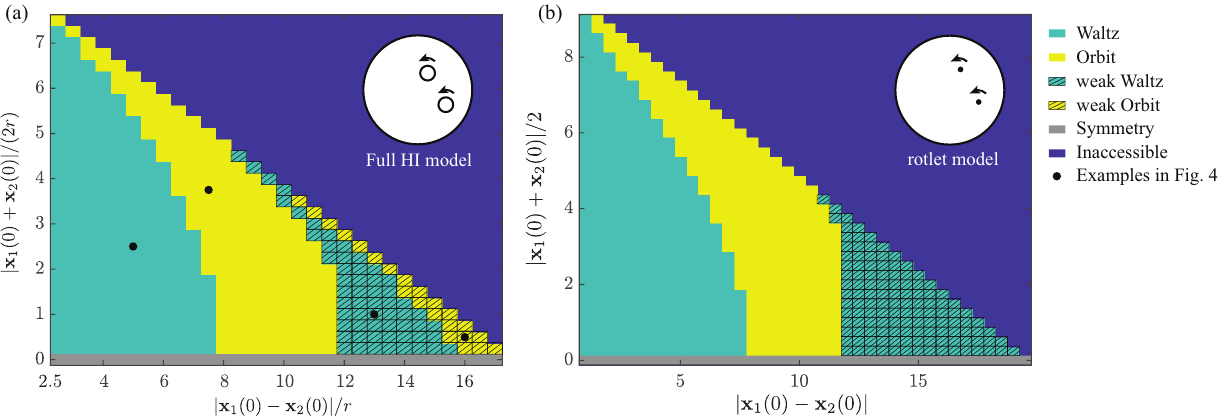}
	\caption[]{Parameter spaces showing distinct hydrodynamic bound states {using the full hydrodynamic interaction model (a) and the rotlet model (b)}. (shaded) Green: (weak) Waltz, (shaded) Yellow: (weak) Orbit, Grey: Symmetry, Dark Blue: Physically inaccessible. Examples shown in Figure~\ref{fig:trajectories} are denoted by black circles.}
	\label{fig:parameterspace}
\end{figure}

Figure~\ref{fig:parameterspace}{(a)} shows the entire parameter space when $\mathbf{x}_1(0)$ and $\mathbf{x}_2(0)$ are varied along the $x$ axis. 
In particular, the horizontal axis is the center-to-center distance within the cylinder pair at $t = 0$, which measures the strength of hydrodynamic interaction between the active cylinders; the vertical axis is the distance of the mid-position of the two active cylinders to the origin at $t = 0$, which measures the eccentricity of the cylinder pair. 
The upper right region of the parameter space is physically inaccessible as at least one active cylinder will be outside the circular confinement.

The cylinder pair is in the equilibrium state by symmetry if $\abs{\mathbf{x}_1(0) + \mathbf{x}_2(0)} = 0$. In this case, both active cylinders will keep moving in the azimuthal direction of the confining cylinder at the same orbital angular velocity.
Aside from the symmetric case, there are
four distinct regions over the entire parameter space, corresponding to the four hydrodynamic bound states discussed earlier in this section.
Specifically, the cylinder pair is in the Waltz bound state if both the initial separation distance $\abs{\mathbf{x}_1(0) - \mathbf{x}_2(0)}$ and the eccentricity are small.
As the separation distance increases, the cylinder pair falls in the Orbit, weak Waltz, and weak Orbit bound states in succession.
For a given initial separation distance, the bound state tends to transition to the next bound state when the eccentricity increases.

{By way of comparison with rotlet models, we adopt the model in \citet{tallapragada2019chaotic} that modeled each micro-roller as a rotlet and present the parameter space in Figure~\ref{fig:parameterspace}(b). 
While the diagrams are largely consistent, our simulation reveals a new hydrodynamic bound state (weak Orbit) that could not be captured using rotlet approximation. This makes sense as the weak Orbit state is only observable when the active cylinder is very close to the bounding surface, in which case the rotlet approximation breaks down.}

\section{Mechanisms of the multiple hydrodynamic bound states}

The motion of each cylinder depends on two factors: (1) The {\em self-induced velocity} resulted from the active torque of its own, and (2) The {\em pair-induced velocity} resulted from the hydrodynamic interaction within the active cylinder pair.
While the self-induced velocity is determined directly by the position of the active cylinder inside the confinement (Fig.~\ref{fig:trans_vel}),
the pair-induced velocity depends both on the distance within the pair ($d_{12}$) and the positions of the active cylinder with respect to the confinement.
In the limiting case where the cylinder pair eccentricity is absent ($\mathbf{x}_1(0) + \mathbf{x}_2(0) = 0$), the system possesses rotational symmetry by construction. Both cylinders will translate in the azimuthal direction of the confinement at the same orbital angular velocity and the pair eccentricity will remain 0 for all $t$, leading to the {\em Symmetry} state shown in Figure~\ref{fig:parameterspace}.
{In the following, we focus on analyzing the four examples shown in Figure~\ref{fig:trajectories} to deduce the mechanisms of these hydrodynamic bound states.}

In example 1, 
the small eccentricity for each active cylinder and the small center-to-center distance within the pair together implies that the self-induced velocity is dominated by the pair-induced velocity.
As a result, the cylinder pair rotates about its center similar to that of the symmetry state, while the pair center slowly orbits around the center of the confinement due to the non-zero eccentricity, resulting in the {\em Waltz} bound state.

In example 2, cylinder 2 is at the center of the secondary recirculation zone generated by the rotation of cylinder 1, at which location both the self- and pair-induced velocities are close to zero. 
Additionally, the flow generated by cylinder 2 advects cylinder 1 in the same direction as its self-induced velocity, reinforcing the azimuthal movement of cylinder 1.
Cylinder 2 is thus trapped close to the origin while cylinder 1 orbits in the azimuthal direction, leading to the {\em Orbit} bound state.

\begin{figure}
\centering
\includegraphics{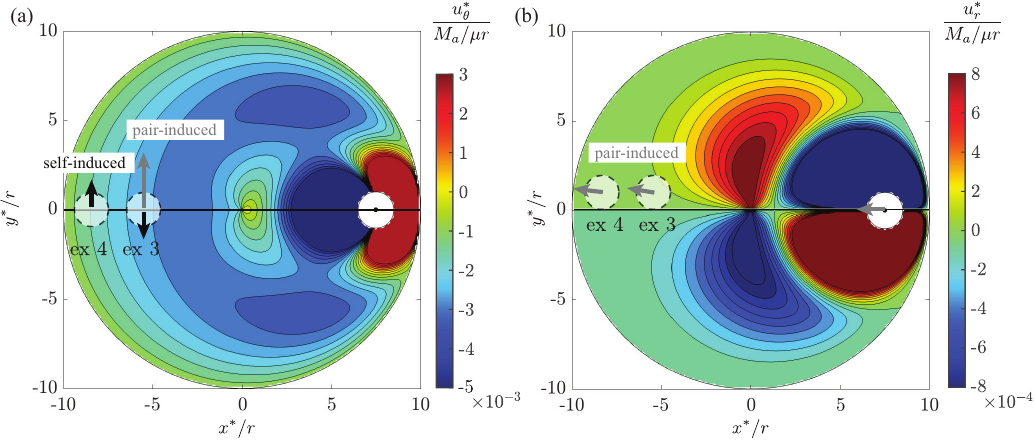}
	\caption[]{Mechanisms for the weak Waltz and the weak Orbit bound states. Panels show the azimuthal (a) and radial (b) velocities generated by the active cylinder at $7.5r\mathbf{e}_x$. The other cylinder's positions leading to the weak Waltz and weak Orbit bound states are denoted by the faint circles. The self- and pair-induced velocities are denoted by black and grey arrows, respectively.}
	\label{fig:mech}
\end{figure}

The analysis of the mechanisms is more complicated for larger center-to-center distances, as the pair-induced velocities no longer dominate the self-induced velocities and the cylinders are in the secondary recirculation zones generated by their counterparts.
To facilitate our analysis, we decompose the velocities generated by the rotation of cylinder 1 into the azimuthal and radial components.
These components are shown in the co-rotating frame as we are mostly focused on the {relative} motions of the pair (Fig.~\ref{fig:mech}).
We note that the radial components of the self-induced velocity as well as that of the pair-induced velocity are both zero if the centers of the cylinders and the origin form a straight line. 
We look at the azimuthal component first.
	
In example 3, the pair-induced azimuthal velocity on cylinder 1 is much lower than its self-induced counterpart because of the wall-screening effect ($3.8\times 10^{-4}M_a/\mu r$ vs $5.3\times 10^{-3} M_a/\mu r$), whereas those on cylinder 2 are comparable to each other ($1.8\times 10^{-3} M_a/\mu r$ vs $4.0\times 10^{-3} M_a/\mu r$).
Therefore, even though the self-induced orbital angular velocity ($\omega_k = U_k/d_k$) of cylinder 2 is slightly higher than that of cylinder 1, as evidenced by the tapered slope of the curve in Figure~\ref{fig:trans_vel}, the pair-induced velocity in the azimuthal direction advects cylinder 2 clockwise to the second quadrant of the co-rotating frame (Fig.~\ref{fig:mech}(a)). 
In the second quadrant, the radial component of the pair-induced velocity is positive, meaning that $d_2$ will increase and catch up $d_1$ (Fig.~\ref{fig:mech}(b)). On the other hand, the flow generated by cylinder 2 in the second quadrant advects cylinder 1 in the negative radial direction,  which initiates the {\em weak Waltz} bound state where the two cylinders will eventually exchange their relative positions.

In example 4, both cylinders are close to the confining boundary with a large center-to-center distance within the pair, thus the self-induced velocities dominate the pair-induced velocities for each cylinder.
The decreasing slope in Figure~\ref{fig:trans_vel} for large $d_k$ shows that the self-induced orbital velocity for cylinder 2 (at $d_2 = 8.5r$) is slower than that for cylinder 1 (at $d_1 = 7.5r$). Therefore cylinder 2 will move into the second quadrant of the co-rotating frame and $d_2$ will increase even though it is already greater than $d_1$ at the beginning. Therefore the two cylinders will not be exchangeable and are in the {\em weak Orbit} bound state.

\section{Conclusions and Discussions}
In this paper, we systematically studied the hydrodynamic bound states of an active cylinder pair rotating inside cylindrical confinement. 
We focused on the case where the two active cylinders are identical both in terms of geometry and active torque. 
We found that the active cylinder pair can fall into four distinct non-trivial hydrodynamic bound states, termed {\em Waltz}, {\em Orbit}, {\em weak Waltz}, and {\em weak Orbit} depending on their initial positions, where Waltz and weak Waltz are the states where the motions of the two active cylinders are exchangeable up to a half-period shift in phase, i.e., $d_1(t) = d_2(t+T/2)$.
The distinction between Waltz and weak Waltz is based on the parameter space, as the two states are separated by the Orbit state. Similarly for Orbit and weak Orbit, where the weak Waltz state separate the two states in the parameter space. 
The mechanisms of these hydrodynamic bound states can be explained by the competing effects of the self-induced velocity generated by each active cylinder and the confinement and the pair-induced velocity generated within the active cylinder pair.

We note that the Waltz bound state is reminiscent of the {\em Leapfrog} motion observed in \citet{delmotte2019hydrodynamically} in which the micro-rollers are placed close to a flat wall.
In fact, the Leapfrog motion is the only periodic state observed in \citep{delmotte2019hydrodynamically} when the micro-rollers are neurally buoyant -- our work shows that having non-planar confinement can lead to various periodic states, highlighting the effects of complex geometry on active matters. 
Some recent work has been investigating the effect of complex geometry on run-and-tumble behaviors~\citep{khatami2016active,forgacs2021active,moen2022trapping,goral2022frustrated}, 
our results suggest that bacteria swimming in cylindrical confinement may exhibit different modes,
as run-and-tumble is essentially the result of rotating (and counter-rotating) a few semi-rigid flagella.

{We focused on the radius ratio $R/r = 10$ in this manuscript. While this large ratio may suggest a possible separation of scales and validate the use of far-field approximations such as the rotlet model, we showed that the rotlet model cannot capture the whole picture, particularly when the active cylinder is close to the confining boundary. On the other hand, qualitatively similar results are obtained with a smaller radius ratio ($R/r = 5$) when the active cylinders are placed along the confinement diameter. It is worth pointing out that smaller radius ratios could lead to more interesting pair dynamics when the cylinders are not placed along the confinement diameter. In fact, we observed another hydrodynamic bound state in which one active cylinder is ``locked behind'' the other cylinder at an approximately constant distance (see supplemental video in \cite{supp}). Systematically studies of this asymmetric effect with different radius ratios require a much bigger parameter space, and are beyond the scope of this paper.}

{Our analytical result of a single active cylinder moving inside the circular confinement was built on the work of \citet{wakiya1975application} that was designed for pinned-cylinder cases. We note that newer approaches using complex methods have been developed and adapted to cylinders with prescribed translational and rotational motions, as shown in \citet{finn2001stokes,finn2003topological}.}

Many extensions can be applied to this work. For example, one can introduce asymmetry within the active cylinder pair, either in terms of the shape or the driving mechanism. This asymmetry could presumably lead to more interesting motions. Furthermore, one can also alter the shape of the confinement and explore the possibility of delivering a cargo cylinder from one compartment to another by controlling the torque of the active rotating cylinder. 
It will also be interesting to see how different bound states affect the mixing of scalar fields~\cite{finn2001stokes,finn2003topological,vikhansky2003chaotic}. 
On the other hand, fast numerical methods such as the Fast multipole method (FMM)~\citep{greengard1987fast,yan2020scalable} can be readily applied to the boundary integral method we adopted here, allowing the possibility to study the interactions of many active cylinders.
Extending the work to 3D would also allow us to study more realistic cases with more interesting geometries such as helical-shape filaments.


\bibliography{references}

\end{document}